%% file: main_SPL.tex
\documentclass[journal]{IEEEtran}

\ifCLASSINFOpdf
\else
   \usepackage[dvips]{graphicx}
\fi

\usepackage[cmex10]{amsmath}
\usepackage{amssymb,amsfonts}
\usepackage{graphicx}
\usepackage{url}
\usepackage{booktabs}
\usepackage{siunitx}
\usepackage[numbers,compress]{natbib}
\usepackage{bm}
\usepackage{array}
\usepackage{mathtools}
\usepackage{multirow}
\usepackage{dblfloatfix}
\usepackage{algorithm}
\usepackage{algpseudocode}

\newcommand{\revise}[1]{#1}

\makeatletter
\let\IEEEorigcaption\caption
\renewcommand{\caption}{\@ifstar{\IEEEcaptionstar}{\IEEEorigcaption}}
\newcommand{\IEEEcaptionstar}[1]{\par\smallskip{\footnotesize #1\par}\smallskip}
\providecommand{\captionsetup}[1]{}
\makeatother
\input{src/command}

\begin{document}
    \input{src/control-bibtex}

    \title{Inner-Loop-Free Total-Variation-Constrained Full-Waveform Inversion}

	\author{
        Shingo~Takemoto,~\IEEEmembership{Member,~IEEE,}
        and Shunsuke~Ono,~\IEEEmembership{Senior Member,~IEEE}





		\input{src/acknowledgement}

	}

    \markboth{IEEE Signal Processing Letters, Vol. XX, No. XX, 2026}
    {Takemoto \MakeLowercase{\textit{et al.}}: Inner-Loop-Free Total-Variation-Constrained Full-Waveform Inversion}

    \maketitle

    \begin{abstract} \input{src/0-summary} \end{abstract}


    \IEEEpeerreviewmaketitle

    \section{Introduction}      \label{sec:Introduction}    \input{src/1-introduction}

    \section{Preliminaries}     \label{sec:Preliminaries}   \input{src/2-preliminaries}

    \section{Proposed Method}   \label{sec:ProposedMethod}  \input{src/3-proposed-method}

    \section{Experiments}       \label{sec:Experiments}     \input{src/4-experiments}

    \section{Conclusion}        \label{sec:Conclusion}      \input{src/5-conclusion}



    \small
    \input{src/main_SPL.bbl}

\end{document}

%% file: src/command.tex
\newcommand{\vecstyle}[1]{#1}
\newcommand{\opstyle}[1]{#1}
\newcommand{\setstyle}[1]{\mathcal{#1}}

\newcommand{\realNum}{\mathbb{R}}
\newcommand{\NumGrid}{N}
\newcommand{\NumData}{M}

\DeclarePairedDelimiter{\norm}{\lVert}{\rVert}

\DeclarePairedDelimiterX{\ip}[2]{\langle}{\rangle}{#1, #2}
\DeclareMathOperator{\prox}{prox}
\DeclareMathOperator{\TV}{TV}
\DeclareMathOperator{\RMSE}{RMSE}
\DeclareMathOperator*{\argmin}{argmin}

\newcommand{\conj}[1]{#1^{*}}
\newcommand{\adj}[1]{#1^{*}}
\newcommand{\IndFunc}[2]{\iota_{#1}(#2)}
\newcommand{\Proj}[1]{P_{#1}}

\newcommand{\vecx}{\vecstyle{x}}
\newcommand{\vecy}{\vecstyle{y}}
\newcommand{\velModel}{\vecstyle{m}}
\newcommand{\seisData}{\vecstyle{u}}
\newcommand{\vecBeta}{\vecstyle{\beta}}

\newcommand{\LinOp}{\opstyle{L}}
\newcommand{\DiffOp}{\opstyle{D}}
\newcommand{\DiffOpHori}{\DiffOp_h}
\newcommand{\DiffOpVert}{\DiffOp_v}

\newcommand{\BoxBall}{\setstyle{B}_{\mathrm{box}}}
\newcommand{\LOneTwoBall}{\setstyle{B}_{\ell_{1,2}}}

%% file: src/control-bibtex.tex
\makeatletter
\def\bstctlcite{\@ifnextchar[{\@bstctlcite}{\@bstctlcite[@auxout]}}
\def\@bstctlcite[#1]#2{\@bsphack
\@for\@citeb:=#2\do{%
\edef\@citeb{\expandafter\@firstofone\@citeb}%
\if@filesw\immediate\write\csname #1\endcsname{\string\citation{\@citeb}}\fi}%
\@esphack}
\makeatother
\bstctlcite{IEEEexample:BSTcontrol}

%% file: src/acknowledgement.tex
\thanks{
    Manuscript received XXX, XXX; revised XXX XXX, XXX.
}
\thanks{
    This work was supported by JST FOREST Grant Number JPMJFR232M, JST CREST Grant Number JPMJCR25Q5, and JSPS KAKENHI Grant Numbers 24K22291, 25H01296, 25K03136, and 26H02536.
}
\thanks{
    S. Takemoto is with the School of Computing, Institute of Science Tokyo, Yokohama, 226-8501, Japan (E-mail: takemoto.s.e14e@m.isct.ac.jp).
}
\thanks{
    S. Ono is with the School of Computing, Institute of Science Tokyo, Yokohama, 226-8501, Japan (E-mail: ono@comp.isct.ac.jp).
}

%% file: src/0-summary.tex
This paper proposes a computationally efficient algorithm to address the Full-Waveform Inversion (FWI) problem with a Total Variation (TV) constraint, designed to accurately reconstruct subsurface properties from seismic signal.
FWI, as an ill-posed inverse problem, requires effective regularizations or constraints to ensure accurate and stable solutions.
Among these, the TV constraint is widely known as a powerful prior for modeling the piecewise smooth structure of subsurface properties.
However, solving the optimization problem is challenging because of the nonlinear observation process combined with the non-smoothness of the TV constraint.
Conventional methods rely on inner loops and/or approximations, which lead to high computational cost and/or inappropriate solutions.
To address these limitations, we develop a novel algorithm based on a primal-dual splitting method, achieving computational efficiency by eliminating inner loops and ensuring high accuracy by avoiding approximations.
We also demonstrate the effectiveness of the proposed method through experiments using the SEG/EAGE Salt Model.
The source code will be available at https://www.mdi.comp.isct.ac.jp/publications/fwiwtv.

%% file: src/1-introduction.tex
\IEEEPARstart{F}{ull-Waveform} Inversion (FWI)~\cite{FWI0,FWI1,FWI2,Zhu2022NNFWI} is a technique for reconstructing subsurface properties from seismic signal measured at multiple observation points.
The subsurface properties obtained through FWI are essential in geological research and resource exploration, such as reservoir characterization and groundwater assessment~\cite{FWI1,FWIApplicationGroundwater0,FWIApplicationGroundwater1}.
FWI has also been applied to non-destructive testing, such as medical imaging and industrial inspection~\cite{FWIApplicationNonDestructiveTesting0,FWIApplicationNonDestructiveTesting1}.


While some seismic analysis tasks focus on restoring the observed seismic signals themselves~\cite{Seismic0,Seismic1,Seismic2}, FWI reconstructs an entire subsurface property model from these signals. 
Since the observed signals depend nonlinearly on the subsurface properties through wave propagation, FWI is typically formulated as an optimization problem that minimizes the squared error between the observed and simulated signals~\cite{FWI0,CustomFWI0,CustomFWI1,CustomFWI2,CustomFWI3,CustomFWI4,CustomFWI5}.
\revise{To accelerate the repeated wave simulations required in this optimization, a recent method replaces the numerical wave-equation solver with a trained neural operator~\cite{Ambient2025Zou}}.
Nevertheless, the inherent ill-posedness of FWI still necessitates regularization techniques.
Among these, Tikhonov regularization~\cite{tikhonov} and Total Variation (TV)-based methods~\cite{TV,TGV} are widely used to promote piecewise smoothness in the reconstructed subsurface properties, improving stability and accuracy~\cite{FWI-with-tikhonov-regularization,FWI-with-TV-regularization,FWI-with-directional-TV-regularization,FWI-with-high-order-TV-regularization,FWI-with-TGV-regularization}.
However, these techniques require tuning a balance parameter that determines the trade-off between the FWI objective function and the regularization term.
To avoid this tuning, several studies incorporate the TV prior as a constraint rather than as a regularization term~\cite{FWI-with-TV-constraint,FWI-with-TV-constraint2,FWI-with-TV-constraint3,FWI-with-TV-constraint4}.
This formulation decouples the parameter for the TV constraint from the FWI objective, allowing it to be determined independently based on prior knowledge of subsurface properties~\cite{constraint0,constraint1,constraint2,constraint3,constraint4,constraints-vs-penalties-in-FWI}.
By doing so, this approach not only simplifies parameter selection but also enhances the interpretability of both the mathematical formulation and the reconstructed subsurface models.

Despite its advantages, solving the TV-constrained FWI problem presents considerable difficulties due to the interplay between the nonlinear observation process and the non-smoothness of the TV constraint.
\revise{Conventional methods attempt to address these issues by iteratively solving constrained subproblems to enforce the constraint at each optimization step~\cite{FWI-with-TV-constraint,FWI-with-TV-constraint3,FWI-with-TV-constraint4} or by employing linear or quadratic approximations~\cite{FWI-with-TV-constraint,FWI-with-TV-constraint2}.}
However, these approaches come with notable drawbacks: inner loops substantially increase computational cost, and approximations compromise reconstruction accuracy.
This raises a crucial question: \textit{Is it possible to develop an algorithm that solves the TV-constrained FWI problem efficiently while avoiding inner loops and approximations?}

In this paper, we introduce a novel algorithm for solving the TV-constrained FWI problem using a primal-dual splitting method~\cite{PDS2}.
The proposed algorithm effectively addresses the intertwined issues of the nonlinear observation process and the non-smoothness of the TV constraint, achieving accurate reconstructions without relying on approximations.
\revise{Additionally, this inner-loop-free design removes the iterative constraint-handling overhead at each FWI iteration.}
We validate the performance of our algorithm through numerical experiments on the SEG/EAGE Salt Model, demonstrating its capability to efficiently enforce the constraints while delivering accurate reconstructions.

%% file: src/2-preliminaries.tex
\subsection{Mathematical Tools} \label{subsec:mathematical-tools}

Throughout this paper, we denote vectors and matrices by lowercase letters (e.g., $\vecx$) and uppercase letters (e.g., $\opstyle{X}$), respectively.

\revise{For $\vecx \in \realNum^{\NumGrid}$, the total variation (TV)~\cite{TV} is defined as
\begin{equation} \label{eq:TVDefinitionEq}
    \TV(\vecx) \coloneqq \norm{\DiffOp\vecx}_{1,2},
\end{equation}
where $\DiffOp \coloneqq [\DiffOpHori^{\top}\ \DiffOpVert^{\top}]^{\top} \in \realNum^{2\NumGrid \times \NumGrid}$, and $\DiffOpHori, \DiffOpVert \in \realNum^{\NumGrid \times \NumGrid}$ are the horizontal and vertical forward-difference operators with Neumann boundary conditions, respectively.
The mixed $\ell_{1,2}$ norm groups the horizontal and vertical differences at the same pixel.}

\subsection{Proximal Tools} \label{subsec:proximal-tools}

We denote the set of proper lower-semicontinuous convex functions $\realNum^{\NumGrid} \to (-\infty,\infty]$ by $\Gamma_0(\realNum^{\NumGrid})$.
For $\gamma > 0$ and $f \in \Gamma_0(\realNum^{\NumGrid})$, the proximity operator is defined as follows:
\begin{equation} \label{eq:ProximityOperatorDefinitionEq}
    \prox_{\gamma f}(\vecx)
    \coloneqq
    \argmin_{\vecy \in \realNum^{\NumGrid}}
    \left\{f(\vecy)+\frac{1}{2\gamma}\norm{\vecy-\vecx}_2^2\right\}.
\end{equation}

For a nonempty closed convex set $C \subset \realNum^{\NumGrid}$, the indicator function $\iota_C: \realNum^{\NumGrid} \to (-\infty,\infty]$ is defined as follows:
\begin{equation} \label{eq:IndicatorFunctionDefinitionEq}
    \IndFunc{C}{\vecx}
    \coloneqq
    \begin{cases}
        0,      & \text{if } \vecx \in C, \\
        \infty, & \text{otherwise}.
    \end{cases}
\end{equation}

The proximity operator of $\iota_C$ is equivalent to the projection onto $C$, given by
\vspace{-1.4mm}
\begin{equation} \label{eq:ProximityOperatorDefinitionWithIndicatorFunctionEq}
    \prox_{\gamma\iota_C}(\vecx)
    = \Proj{C}(\vecx)
    \coloneqq \argmin_{\vecy \in C}\norm{\vecy-\vecx}_2.
\end{equation}

\subsection{Primal-Dual Splitting Algorithm} \label{subsec:primal-dual-splitting-algorithm}

The Primal-Dual Splitting algorithm (PDS)~\cite{PDS2} is applied to the following problem:
\begin{equation} \label{eq:PDSPrimalEq}
    \min_{\vecx \in \realNum^{\NumGrid}}
    \left\{f(\vecx)+g(\vecx)+h(\LinOp\vecx)\right\},
\end{equation}
where $\LinOp \in \realNum^{\NumData \times \NumGrid}$ is a linear operator, $f$ is a differentiable convex function and $g,h$ are convex functions whose proximity operators can be computed efficiently.

PDS solves Prob.~\eqref{eq:PDSPrimalEq} by iteratively updating the following:
\begin{equation} \label{eq:PDSSubStep}
\left\lfloor
    \begin{aligned}
        \vecx^{(k+1)}
        &= \prox_{\gamma_1g}\!\left(
            \vecx^{(k)}
            -\gamma_1\bigl(
                \nabla f(\vecx^{(k)})
                +\adj{\LinOp}\vecy^{(k)}
            \bigr)
        \right), \\
        \vecy^{(k+1)}
        &= \prox_{\gamma_2\conj{h}}\!\left(
            \vecy^{(k)}
            +\gamma_2\LinOp
            \bigl(2\vecx^{(k+1)}-\vecx^{(k)}\bigr)
        \right).
    \end{aligned}
\right.
\end{equation}
where $\conj{h}$ is the \emph{Fenchel--Rockafellar conjugate function}\footnote{The Fenchel--Rockafellar conjugate function of $h$ is defined by $\conj{h}(\bm{z}) \coloneqq \sup_{\bm{v} \in \realNum^{\NumData}}\{\ip{\bm{v}}{\bm{z}}-h(\bm{v})\}$. The proximity operator of $\conj{h}$ can be expressed as $\prox_{\gamma\conj{h}}(\bm{z})=\bm{z}-\gamma\prox_{h/\gamma}(\bm{z}/\gamma)$~\cite[Theorem 3.1(ii)]{prox-convex-conjugate-function}.} of $h$, and $\gamma_1, \gamma_2 > 0$ are step sizes.
Under appropriate step-size conditions, $\vecx^{(k)}$ converges to a solution of \eqref{eq:PDSPrimalEq}~\cite{PDS2}.

\subsection{Full-Waveform Inversion (FWI)} \label{subsec:full-waveform-inversion}

Typically, FWI is treated as the following optimization problem~\cite{FWI1}:
\begin{equation} \label{eq:FWIObjective}
    \argmin_{\velModel \in \realNum^{\NumGrid}}
    E(\velModel)
    =
    \frac{1}{2}
    \norm{\seisData_{\mathrm{obs}}-\seisData_{\mathrm{cal}}(\velModel)}_2^2,
\end{equation}
where $\velModel \in \realNum^{\NumGrid}$ is the velocity model representing subsurface properties and
$\seisData_{\mathrm{obs}} \in \realNum^{\NumData}$ is the observed seismic signal.
$\NumGrid$ is the number of grid points, and $\NumData$ is the total data size of the observed seismic signal, defined as the total product of the number of waveform sources, time samples, and receivers.
\revise{The observation process $\seisData_{\mathrm{cal}}(\velModel)$ is defined through wave propagation for each source~\cite{FWI0}. Specifically, for each source $s$, let $\bm{p}_s(\velModel)$ denote the wavefield satisfying
\begin{equation} \label{eq:FWIWaveEquation}
    \bm{A}(\velModel)\bm{p}_s(\velModel)=\bm{q}_s,
\end{equation}
where $\bm{A}(\velModel)$ is the discretized wave-equation operator and $\bm{q}_s$ is the $s$-th source. 
Then, 
$\seisData_{\mathrm{cal}}(\velModel)
=\bigl[\bigl(\bm{R}\bm{p}_1(\velModel)\bigr)^{\top},
\ldots,\bigl(\bm{R}\bm{p}_S(\velModel)\bigr)^{\top}\bigr]^{\top}$,
where $\bm{R}$ extracts the wavefield values at the receiver locations over all time samples and $S$ is the number of waveform sources.
Since $\bm{p}_s(\velModel)$ depends nonlinearly on $\velModel$, $\seisData_{\mathrm{cal}}(\velModel)$ is also nonlinear.
Consequently, $E$ is generally nonconvex.
However, the gradient $\nabla E(\velModel)$ can be efficiently computed using the adjoint-state method~\cite{FWI-gradient}.}



%% file: src/3-proposed-method.tex

We introduce the TV and box constraints into the FWI problem to achieve more accurate reconstruction.
As shown in the ground-truth velocity model in the top-left panel of Fig.~\ref{fig:velocity-models-combined}, the velocity model is piecewise smooth, thus introducing the TV constraint to achieve a more accurate reconstruction.
The box constraint ensures that the velocity model remains within valid ranges.

The optimization problem of the TV and box constrained FWI is formulated as follows:
\begin{equation} \label{eq:FWIObjectiveWithTVConstraint}
    \argmin_{\velModel \in \realNum^{\NumGrid}} E(\velModel)
    \quad \text{s.t.} \quad
    \begin{cases}
        \norm{\DiffOp\velModel}_{1,2} \leq \alpha, \\
        \velModel \in [l,u]^{\NumGrid},
    \end{cases}
\end{equation}
where $\alpha \ge 0$ is the upper bound of the $\ell_{1,2}$ norm, and $u > l > 0$ are the upper and lower bounds of the velocity model values, respectively.
As highlighted in~\cite{constraint0,constraint1,constraint2,constraint3,constraint4,constraints-vs-penalties-in-FWI}, this constrained formulation allows us to determine $\alpha$ independently of the other terms and to control the smoothness directly.

The constraints can be incorporated into the objective function as indicator functions:
\begin{equation} \label{eq:FWIObjectiveWithTVConstraintWithIndicatorFunction}
    \argmin_{\velModel \in \realNum^{\NumGrid}}
    \left\{
        E(\velModel)
        +\IndFunc{\BoxBall}{\velModel}
        +\IndFunc{\LOneTwoBall}{\DiffOp\velModel}
    \right\},
\end{equation}
where
\begin{equation}
    \label{eq:LOneTwoBallDefinition}
    \BoxBall \coloneqq [l,u]^{\NumGrid}, \quad
    \LOneTwoBall \coloneqq \left\{\norm{\cdot}_{1,2}\leq\alpha\right\}.
\end{equation}

Prob.~\eqref{eq:FWIObjectiveWithTVConstraintWithIndicatorFunction} corresponds to Prob.~\eqref{eq:PDSPrimalEq} with $\vecx=\velModel$, $f=E$, $g=\iota_{\BoxBall}$, $h=\iota_{\LOneTwoBall}$, and $\LinOp=\DiffOp$.
\revise{However, since $E$ is generally nonconvex, the convergence guarantee of PDS does not apply to the proposed algorithm.
The adjoint-state method is used only to evaluate $\nabla E$, whereas the TV and box constraints are handled separately through the primal-dual projection steps; these constraints do not eliminate the nonconvexity of $E$.
Hence, as in conventional FWI, the lack of low-frequency components can aggravate cycle skipping in the proposed method~\cite{FWI1}.}
We show the detailed algorithm in Algorithm~\ref{alg:FWIWithPDS}.
Applying the PDS updates to~\eqref{eq:FWIObjectiveWithTVConstraintWithIndicatorFunction} yields an inner-loop-free algorithm in which the TV and box constraints are handled without approximations or relaxations.

\begin{algorithm}[t]
    \caption{PDS based solver for~\eqref{eq:FWIObjectiveWithTVConstraintWithIndicatorFunction}}\label{alg:FWIWithPDS}
    \begin{algorithmic}[1]
        \Statex \textbf{Input:} $\velModel^{(0)}, \vecy^{(0)}, \gamma_1 > 0, \gamma_2 > 0$
        \While {A stopping criterion is not satisfied}
            \State $\widetilde{\velModel} \leftarrow
                \velModel^{(k)}
                -\gamma_1\left(
                    \nabla E(\velModel^{(k)})
                    +\adj{\DiffOp}\vecy^{(k)}
                \right)$
            \State $\velModel^{(k+1)} \leftarrow
                \Proj{\BoxBall}(\widetilde{\velModel})$
            \State $\widetilde{\vecy} \leftarrow
                \vecy^{(k)}
                +\gamma_2\DiffOp
                \left(2\velModel^{(k+1)}-\velModel^{(k)}\right)$
            \State $\vecy^{(k+1)} \leftarrow
                \widetilde{\vecy}
                -\gamma_2\Proj{\LOneTwoBall}
                \left(\widetilde{\vecy}/\gamma_2\right)$
        \EndWhile
        \Statex \textbf{Output:} $\velModel^{(k)}$
    \end{algorithmic}
\end{algorithm}

The projections onto $\BoxBall$ and $\LOneTwoBall$ are given by
\begin{equation} \label{eq:ProximityOperatorForBoxConstraint}
    \Proj{\BoxBall}(\vecx)
    = \min\!\left(\max(\vecx,l),u\right),
\end{equation}
\begin{equation} \label{eq:ProximityOperatorForL12Ball}
    \left(\Proj{\LOneTwoBall}(\vecx)\right)_{\mathfrak{g}_i}
    =
    \begin{cases}
        0,
        & \text{if } \norm{\vecx_{\mathfrak{g}_i}}_2=0, \\
        \displaystyle
        \vecBeta_i\frac{\vecx_{\mathfrak{g}_i}}{\norm{\vecx_{\mathfrak{g}_i}}_2},
        & \text{otherwise},
    \end{cases}
\end{equation}
where
\begin{equation} \label{eq:ProximityOperatorForL12BallWhere}
    \vecBeta
    =
    \Proj{\{\norm{\cdot}_1\leq\alpha\}}
    \left(
        \left[
            \norm{\vecx_{\mathfrak{g}_1}}_2,
            \ldots,
            \norm{\vecx_{\mathfrak{g}_{\NumGrid}}}_2
        \right]^{\top}
    \right),
\end{equation}
and $\mathfrak{g}_i$ is an index set corresponding to the horizontal and vertical differences of the $i$-th element of $\velModel$.
\revise{In the projection problem onto $\LOneTwoBall$, if the norm of each group of the solution is fixed to $\vecBeta_i$, the optimal group is $\vecBeta_i\vecx_{\mathfrak{g}_i}/\norm{\vecx_{\mathfrak{g}_i}}_2$, which has the same direction as $\vecx_{\mathfrak{g}_i}$.
Hence, the problem reduces to the projection of the group-magnitude vector onto the $\ell_1$-norm ball in~\eqref{eq:ProximityOperatorForL12BallWhere}, which can be efficiently computed using the method of Duchi et al.~\cite{L1-ball-projection}.}


\revise{
\newtheorem{theorem}{Remark}
\input{src/remark-computational-cost}
}

%% file: src/remark-computational-cost.tex
\begin{theorem}[Computational Cost of Our Algorithm]
In the proposed algorithm, the mixed-$\ell_{1,2}$-ball projection defined by \eqref{eq:ProximityOperatorForL12Ball}--\eqref{eq:ProximityOperatorForL12BallWhere} is computed in $O(\NumGrid\log\NumGrid)$ without inner loops.
In contrast, several existing constrained-FWI methods~\cite{FWI-with-TV-constraint, FWI-with-TV-constraint3, FWI-with-TV-constraint4} require iterative inner solvers for constraint handling.
For example, Esser et al.~\cite{FWI-with-TV-constraint} iteratively solve a constrained convex subproblem until convergence, Yong et al.~\cite{FWI-with-TV-constraint3} use up to 50 inner PDHG iterations per FWI update, and Peters et al.~\cite{FWI-with-TV-constraint4} use Dykstra's algorithm to project onto the intersection of multiple constraints.
Hence, the constraint-handling cost of these methods grows with the number of inner iterations, whereas that of the proposed algorithm remains $O(N\log N)$ per iteration.
Consequently, the computational cost of each iteration in the proposed algorithm is dominated by the gradient computation $\nabla E$, which involves wave-equation simulations and has a complexity of $O(ST\NumGrid)$, where $T$ is the number of time samples.
\end{theorem}

%% file: src/4-experiments.tex
\subsection{Experimental Setup} \label{subsec:experimental-setup}

To demonstrate the effectiveness of the TV and box constrained FWI, we conducted FWI experiments where we compared with the standard FWI method\footnote{
    \begin{minipage}{0.95\linewidth}
        The standard FWI method uses the following procedures:
        \begin{equation}
            \velModel^{(k+1)}
            =
            \velModel^{(k)}
            -\gamma\nabla E(\velModel^{(k)})
            \label{eq:FWIWithGradient}
        \end{equation}
        where $\gamma > 0$ is the step size.
    \end{minipage}
}\cite{FWI0} 
and the neural-network-based FWI (NN-FWI) method~\cite{Zhu2022NNFWI}, which retains the wave-equation solver and introduces a prior on the velocity model through neural-network reparameterization.

We use the SEG/EAGE Salt Model with $50 \times 100$ grid points.
The top-left and bottom-left panels of Fig.~\ref{fig:velocity-models-combined} show the ground truth velocity model and the initial velocity model, respectively. The initial velocity model is created by smoothing the ground truth with a Gaussian filter with a standard deviation of 80.
The source waveform is a Ricker wavelet with a peak wavelet frequency of 10 Hz.
The number of waveform sources and receivers is 20 and 101, respectively, and they are placed on the surface at equal intervals.
The gradient $\nabla E$ is computed numerically using the Devito framework~\cite{devito}.
The number of iterations is set to 5000.
Experiments are conducted with and without noise in the observed data, as shown in Fig.~\ref{fig:observed-seismic-data}.
The noise is Gaussian noise with mean 0 and variance 1.
In our algorithm, the step sizes $\gamma_1$ and $\gamma_2$ are set to $1.0 \times 10^{-4}$ and $1.0 \times 10^2$, respectively.
We set $l=1.5$~km/s and $u=4.5$~km/s, and varied $\alpha$ from 100 to 700 in steps of 50.
In the standard FWI method, the step size $\gamma$ is set to $1.0 \times 10^{-4}$.

For quality measures, we employ the root-mean-square error (RMSE) [m/s], defined as $\RMSE(\velModel,\velModel_{\mathrm{true}}) = \sqrt{\frac{1}{\NumGrid}\norm{\velModel-\velModel_{\mathrm{true}}}_2^2}$, and the structural similarity index measure (SSIM)~\cite{SSIM2004Wang}, where $\velModel$ and $\velModel_{\mathrm{true}}$ are the reconstructed and ground-truth velocity models [m/s], respectively.
\revise{For SSIM, the data range is fixed at $u-l=(4.5-1.5)=3.0\,\mathrm{km/s}$ for all methods.}

\input{src/img/observed-seismic-data}
\input{src/img/experiment-results-combined-2x6}
\input{src/img/ssim-graph}

\subsection{Results and Discussion} \label{subsec:results-and-discussion}
Fig.~\ref{fig:velocity-models-combined} shows the reconstructed velocity models with their RMSE and SSIM. The best parameter of the proposed method is $\alpha = 350$, where $\alpha = 150$ represents a stronger TV constraint and $\alpha = 550$ represents a weaker one.
For the quantitative results, NN-FWI achieves the highest SSIM in the noiseless case. However, its performance significantly deteriorates in the noisy case. In the other cases, the proposed method with $\alpha = 350$ achieves the best RMSE and SSIM. 

We next discuss the visual quality of the reconstructed velocity models. NN-FWI reconstructs the main structure to some extent in the noiseless case, but its reconstruction quality substantially degrades in the noisy case. By contrast, the standard FWI method still reconstructs the overall structure even in the noisy case. This is because $E(\velModel)$ minimizes the misfit rather than enforcing $\seisData_{\mathrm{cal}}(\velModel) = \seisData_{\mathrm{obs}}$, which makes the reconstruction less sensitive to the observation noise. However, the standard FWI method generates wave-like artifacts in both the noiseless and noisy cases. In contrast, the proposed method with $\alpha = 350$ reconstructs the velocity model without such artifacts in both cases. This indicates that solving the optimization problem with the two constraints exactly incorporated yields accurate and robust reconstruction.


To analyze the iteration behavior and the effect of the TV constraint parameter, we focus on the standard FWI method and the proposed method.
Fig.~\ref{fig:iters-ssim-noisy} shows the SSIM against the number of iterations for our proposed method and the standard FWI method.
The standard FWI method with the noisy observed seismic signal degrades the SSIM after a certain number of iterations.
On the other hand, the proposed method maintains a consistently high value without degrading the SSIM even after a large number of iterations.
This shows that the proposed method effectively reduces the risk of overfitting to noisy data, providing stable and accurate performance even as the number of iterations increases.

For a detailed analysis of the TV constraint parameter $\alpha$, we plot the RMSE and SSIM of our proposed method against the parameter $\alpha$ and the standard FWI method in Fig.~\ref{fig:alpha-rmse} and Fig.~\ref{fig:alpha-ssim}.
When around $\alpha$ = 350, the proposed method achieves the best RMSE and SSIM.
When $\alpha$ is smaller, the TV constraint becomes stronger, the RMSE and SSIM worsen, and over-smoothing occurs, as in the noiseless result for $\alpha$ = 150 in Fig.~\ref{fig:velocity-models-combined}.
On the other hand, when $\alpha$ is larger, the TV constraint becomes weaker and the result is similar to the standard FWI method, as in the noiseless result for $\alpha$ = 550 in Fig.~\ref{fig:velocity-models-combined}.
However, thanks to the box constraint, the proposed method still outperforms the standard FWI method.
This demonstrates that the parameter $\alpha$ has a clear and predictable effect on the reconstructed velocity model, which can be easily adjusted to achieve accurate results.

%% file: src/img/observed-seismic-data.tex
\begin{figure}[t]
    \centering
    \begin{tabular}{m{3mm} m{30mm} m{30mm} m{10mm}}
        \begin{minipage}[b]{\linewidth}\end{minipage} &

        \begin{minipage}[b]{\linewidth}
            \centering
            \includegraphics[width=\linewidth]{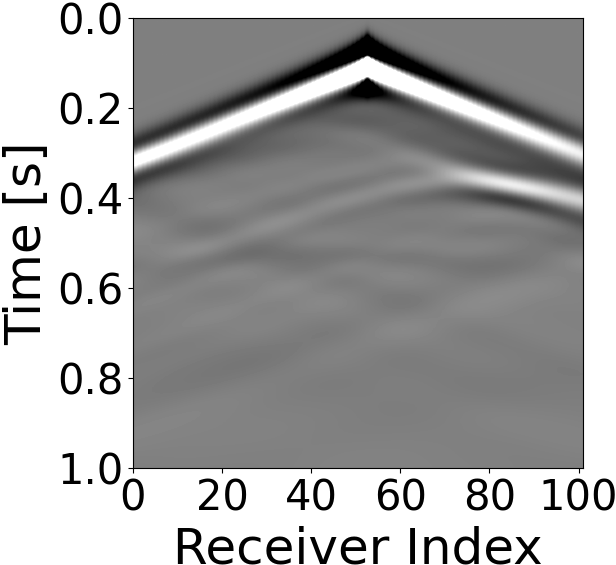}
        \end{minipage} &
        \begin{minipage}[b]{\linewidth}
            \centering
            \includegraphics[width=\linewidth]{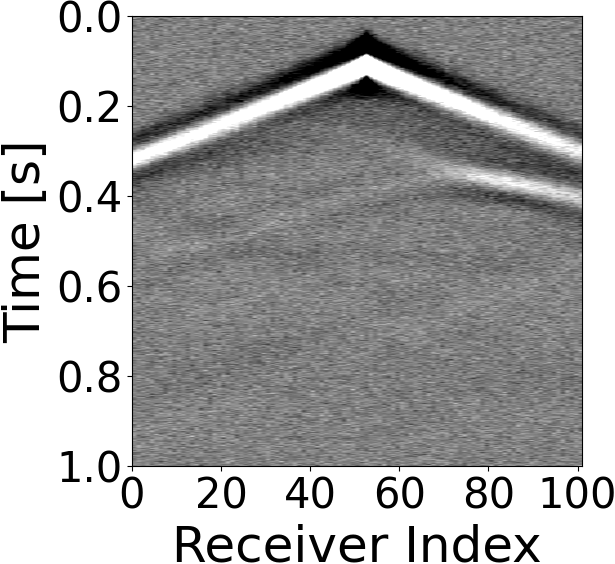}
        \end{minipage} &
        \hspace{-3mm}
        \multirow[t]{3}{*}{\raisebox{-5.8mm}{\includegraphics[height=24mm]{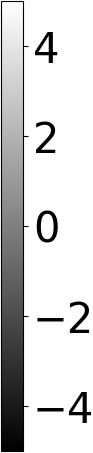}}} \\
    \end{tabular}
    \captionsetup{margin=1.3cm}
    \caption{The synthesized noiseless (left) and noisy (right) seismic signals corresponding to a single source.}
    \label{fig:observed-seismic-data}
\end{figure}

%% file: src/img/experiment-results-combined-2x6.tex
\begin{figure*}[htbp]
    \newcommand{\gapA}{0mm}
    \newcommand{\gapB}{-0.5mm}
    \newcommand{\gapC}{0mm}
    \newcommand{\resultWidth}{28mm}

    \centering
    \begin{tabular}{@{}c@{\hspace{0.8mm}}c@{\hspace{0.8mm}}c@{\hspace{0.8mm}}c@{\hspace{0.8mm}}c@{\hspace{0.8mm}}c@{\hspace{0.8mm}}c@{}}
        \begin{minipage}[b]{\resultWidth}
            \centering
            \includegraphics[width=\linewidth]{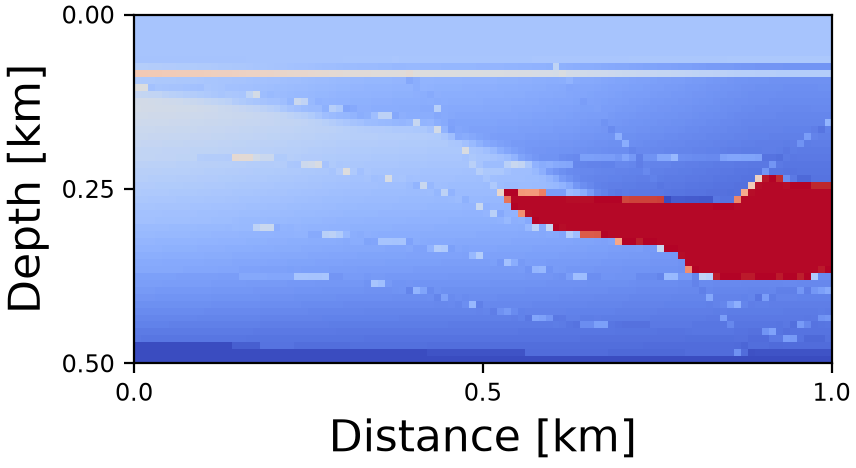}\\[\gapA]
            {\scriptsize RMSE [m/s] \\[\gapB] SSIM} \\[\gapC]
            {\footnotesize Ground-truth}
        \end{minipage} &
        \begin{minipage}[b]{\resultWidth}
            \centering
            \includegraphics[width=\linewidth]{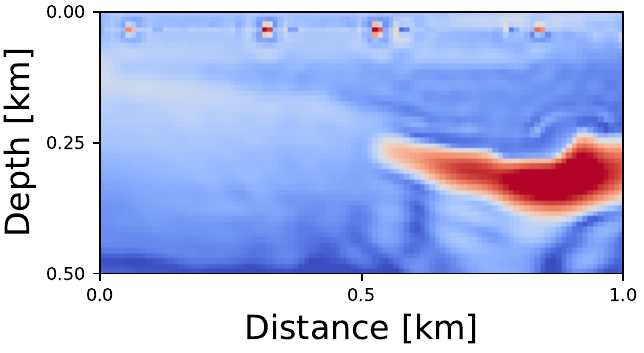}\\[\gapA]
            {\scriptsize 347.80 \\[\gapB] 0.6100} \\[\gapC]
            {\footnotesize Standard FWI}
        \end{minipage} &
        \begin{minipage}[b]{\resultWidth}
            \centering
            \includegraphics[width=\linewidth]{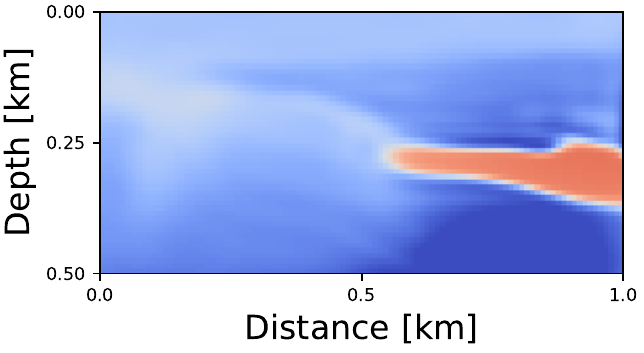}\\[\gapA]
            {\scriptsize 449.81 \\[\gapB] \textbf{0.7706}} \\[\gapC]
            {\footnotesize NN-FWI}
        \end{minipage} &
        \begin{minipage}[b]{\resultWidth}
            \centering
            \includegraphics[width=\linewidth]{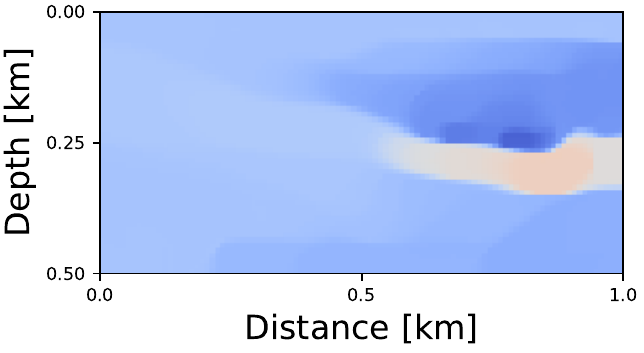}\\[\gapA]
            {\scriptsize 553.77 \\[\gapB] 0.5229} \\[\gapC]
            {\footnotesize Ours $\alpha=150$}
        \end{minipage} &
        \begin{minipage}[b]{\resultWidth}
            \centering
            \includegraphics[width=\linewidth]{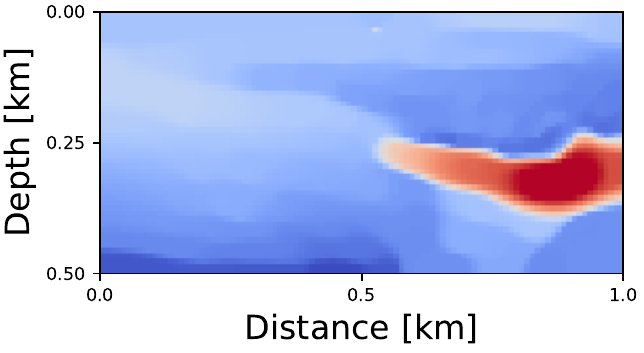}\\[\gapA]
            {\scriptsize \textbf{309.92} \\[\gapB] 0.6707} \\[\gapC]
            {\footnotesize \textbf{Ours $\alpha=350$}}
        \end{minipage} &
        \begin{minipage}[b]{\resultWidth}
            \centering
            \includegraphics[width=\linewidth]{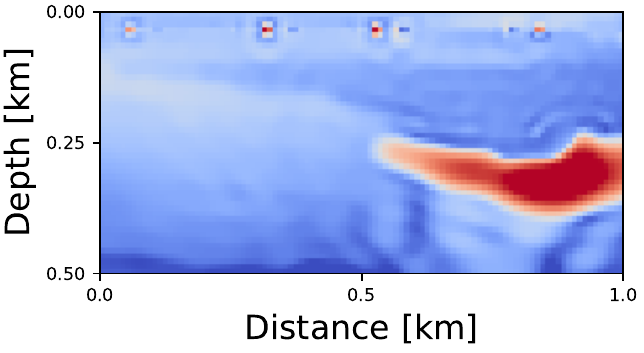}\\[\gapA]
            {\scriptsize 339.92 \\[\gapB] 0.6214} \\[\gapC]
            {\footnotesize Ours $\alpha=550$}
        \end{minipage} & {} \\[1.8mm]

        \begin{minipage}[b]{\resultWidth}
            \centering
            \includegraphics[width=\linewidth]{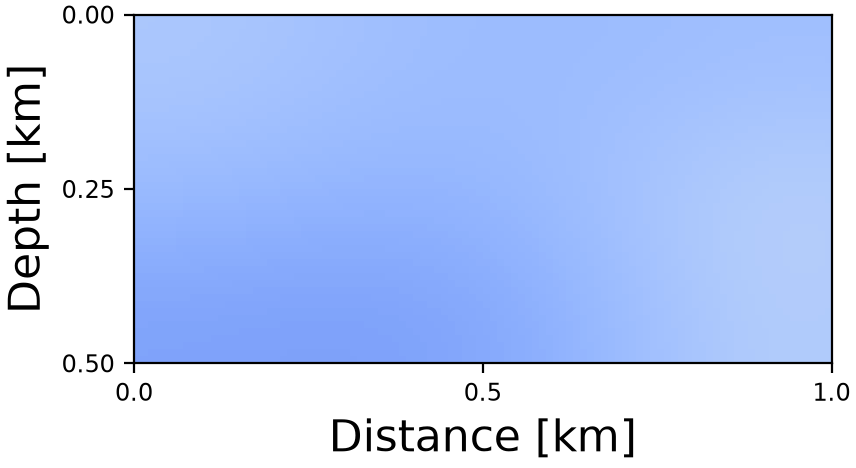}\\[\gapA]
            {\scriptsize RMSE [m/s] \\[\gapB] SSIM} \\[\gapC]
            {\footnotesize Initial model}
        \end{minipage} &
        \begin{minipage}[b]{\resultWidth}
            \centering
            \includegraphics[width=\linewidth]{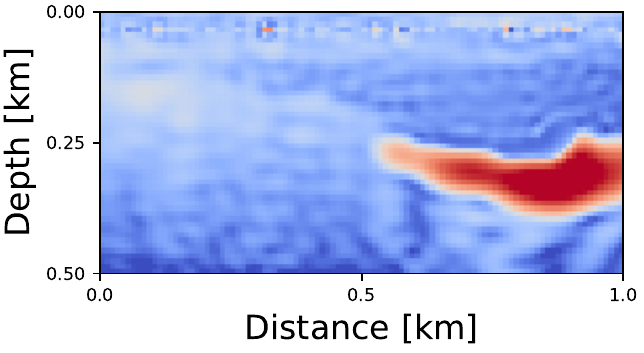}\\[\gapA]
            {\scriptsize 345.68 \\[\gapB] 0.5342} \\[\gapC]
            {\footnotesize Standard FWI}
        \end{minipage} &
        \begin{minipage}[b]{\resultWidth}
            \centering
            \includegraphics[width=\linewidth]{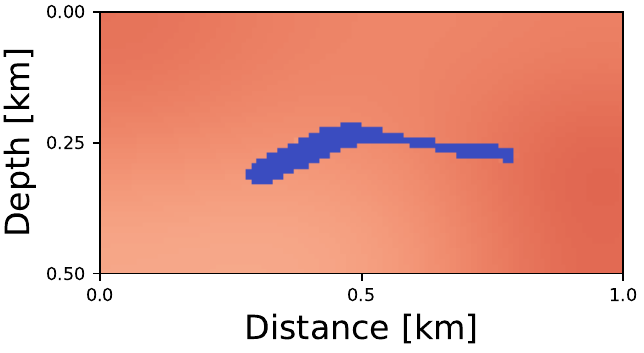}\\[\gapA]
            {\scriptsize 1481.53 \\[\gapB] 0.2616} \\[\gapC]
            {\footnotesize NN-FWI}
        \end{minipage} &
        \begin{minipage}[b]{\resultWidth}
            \centering
            \includegraphics[width=\linewidth]{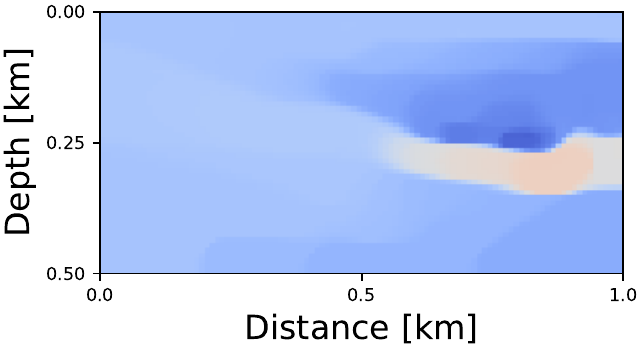}\\[\gapA]
            {\scriptsize 554.59 \\[\gapB] 0.5230} \\[\gapC]
            {\footnotesize Ours $\alpha=150$}
        \end{minipage} &
        \begin{minipage}[b]{\resultWidth}
            \centering
            \includegraphics[width=\linewidth]{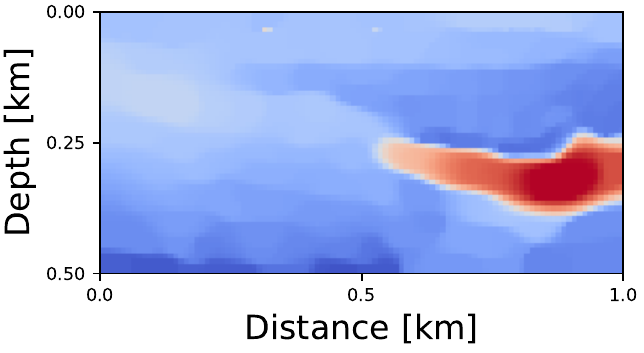}\\[\gapA]
            {\scriptsize \textbf{307.75} \\[\gapB] \textbf{0.6559}} \\[\gapC]
            {\footnotesize \textbf{Ours $\alpha=350$}}
        \end{minipage} &
        \begin{minipage}[b]{\resultWidth}
            \centering
            \includegraphics[width=\linewidth]{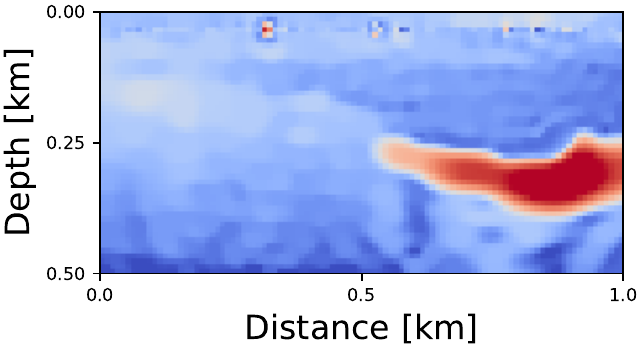}\\[\gapA]
            {\scriptsize 339.33 \\[\gapB] 0.5896} \\[\gapC]
            {\footnotesize Ours $\alpha=550$}
        \end{minipage} &
        \smash{\raisebox{11mm}{\includegraphics[height=32mm]{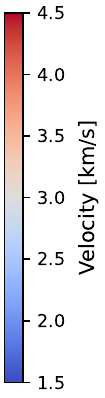}}} \\
    \end{tabular}
    \captionsetup{margin=1cm}
    \caption{
        Velocity models [km/s] and their corresponding reconstructions.
        The leftmost column shows the ground-truth model in the top row and the initial velocity model in the bottom row.
        The remaining columns show the reconstruction results without noise in the top row and with the noisy observed seismic signal in the bottom row.
        The RMSE [m/s] and SSIM are shown below each reconstruction, with the best value for each metric in each case highlighted in bold.
        Among the parameters for the proposed method, $\alpha = 350$ achieves the best performance in both the noiseless and noisy cases.
    }
    \vspace{3mm}
    \label{fig:velocity-models-combined}
\end{figure*}

%% file: src/img/ssim-graph.tex
\begin{figure*}[htbp]
    \centering
    \hspace{-3mm}
    \begin{minipage}{58mm}
        \centering
        \includegraphics[width=\linewidth]{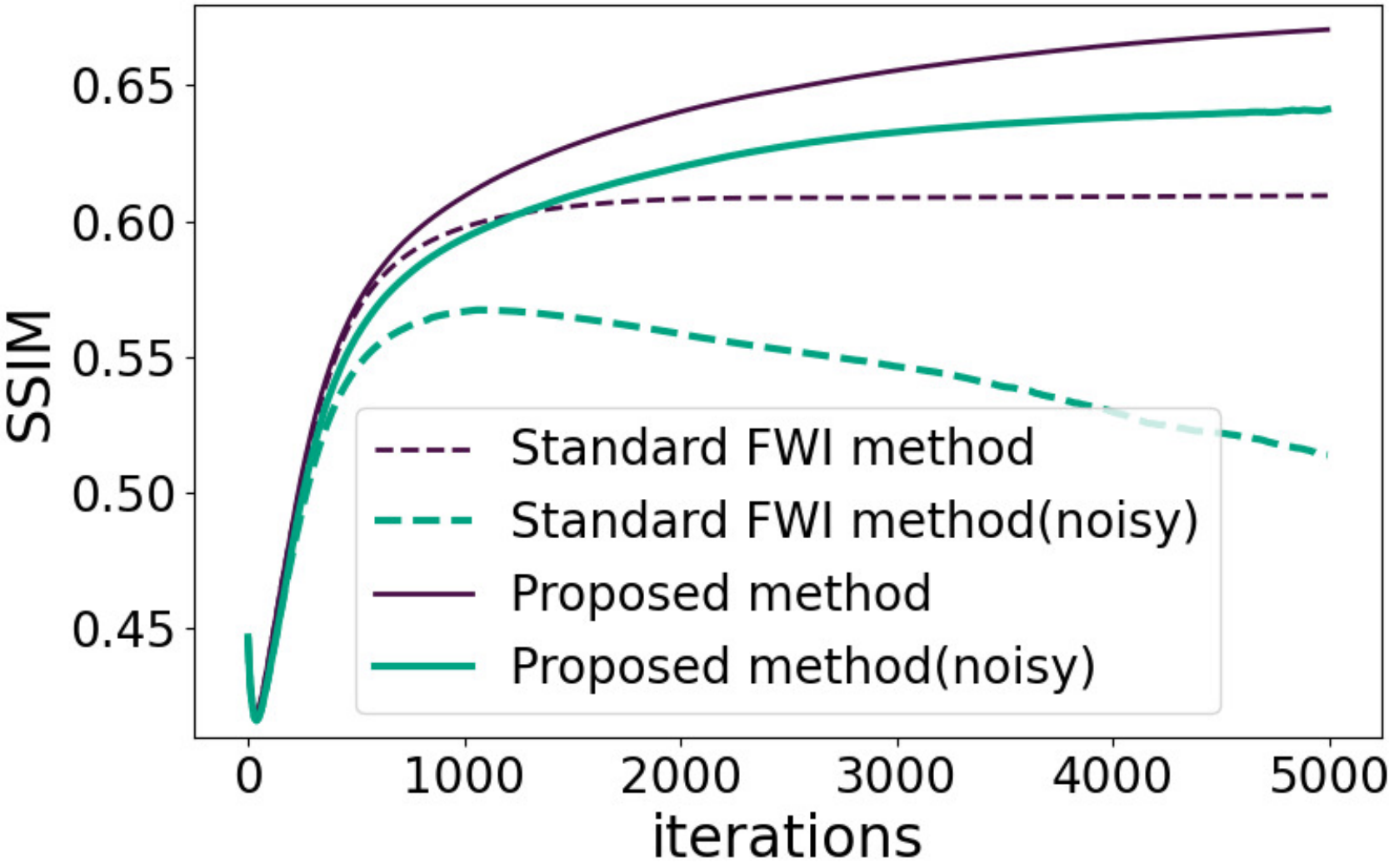}
        \caption{SSIM against iters ($\alpha$ = 350).}
        \label{fig:iters-ssim-noisy}
        \vspace{3mm}
    \end{minipage}
    \hspace{-1mm}
    \begin{minipage}{58mm}
        \centering
        \includegraphics[width=\linewidth]{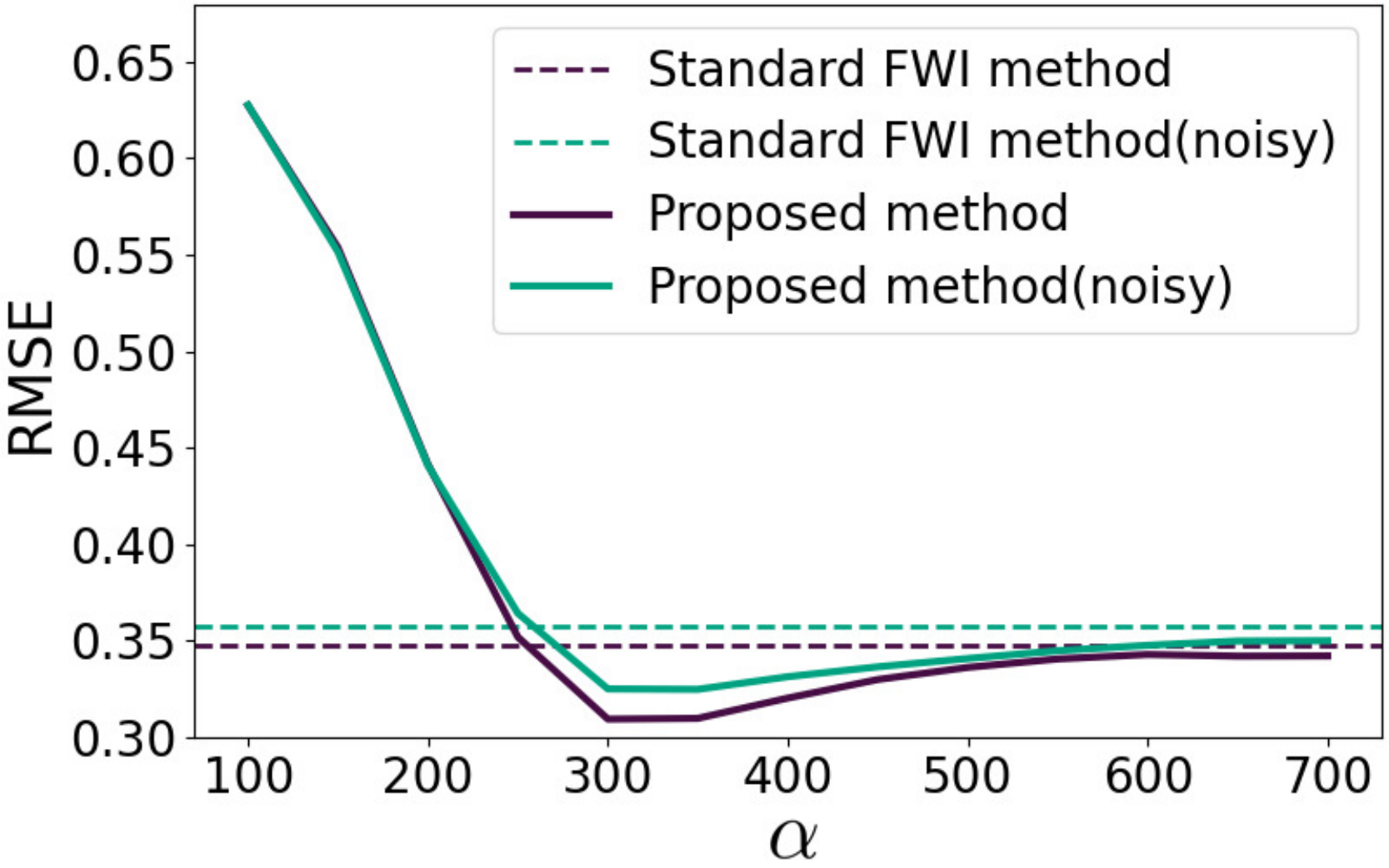}
        \hspace{3mm}
        \begin{minipage}{55mm}
            \caption{RMSE against $\alpha$.}
            \label{fig:alpha-rmse}
        \end{minipage}
        \vspace{3mm}
    \end{minipage}
    \hspace{-2mm}
    \begin{minipage}{58mm}
        \centering
        \includegraphics[width=\linewidth]{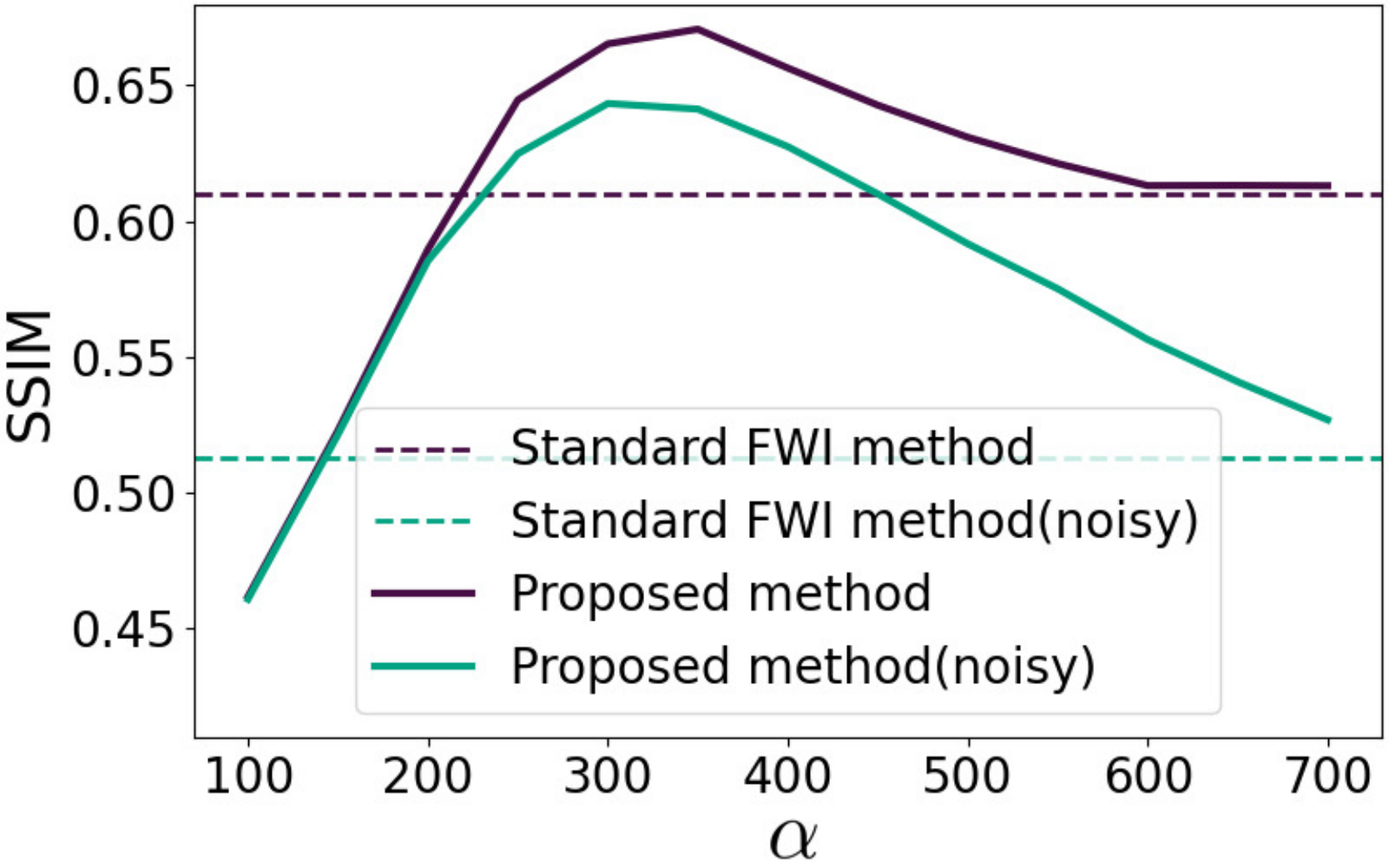}
        \hspace{4mm}
        \begin{minipage}{54mm}
            \caption{SSIM against $\alpha$.}
            \label{fig:alpha-ssim}
        \end{minipage}
    \end{minipage}
\end{figure*}

%% file: src/5-conclusion.tex

\revise{In this paper, we developed an efficient algorithm for TV- and box-constrained FWI based on PDS. The proposed algorithm directly incorporates the TV and box constraints without approximations or iterative inner loops. Experimental results showed that the proposed method suppresses the wave-like artifacts observed in standard FWI and maintains stable reconstruction quality even in the presence of observation noise. The experiments with different values of $\alpha$ further demonstrated the explicit control of the TV constraint over the smoothness of the reconstructed velocity model. To further improve the numerical efficiency of the proposed algorithm, we will investigate adaptive step-size strategies in future work.}